\documentclass[pdflatex,sn-mathphys-num]{sn-jnl}
\usepackage{graphicx}%
\usepackage{multirow}%
\usepackage{amsmath,amssymb,amsfonts}%
\usepackage{amsthm}%
\usepackage{mathrsfs}%
\usepackage[title]{appendix}%
\usepackage{xcolor}%
\usepackage{textcomp}%
\usepackage{manyfoot}%
\usepackage{booktabs}%
\usepackage{algorithm}%
\usepackage{algorithmicx}%
\usepackage{algpseudocode}%
\usepackage{listings}%
\usepackage[utf8]{inputenc} 
\usepackage{gensymb}  
\usepackage{upgreek} 
\usepackage{hyphenat}
\usepackage{placeins}
\usepackage{float} 
\usepackage{subcaption}

\begin{document}
	
	\title[Sub-100 ps TOF Detector with SiPM-Readout Scintillator]{Development of a sub-100 ps Time-of-Flight detector with SiPM-readout scintillator for measurement of cosmic muon velocity}
	
	\author[1]{\fnm{Ziyi} \sur{Yang}}\email{23300200002@m.fudan.edu.cn}
	\author[1]{\fnm{Xiyang} \sur{Wang}}
	\author[1]{\fnm{Shiming} \sur{Zou}}
	\author[1]{\fnm{Ting} \sur{Wang}}
	\author[1]{\fnm{Kairui} \sur{Huang}}
	\author[1]{\fnm{Wanyi} \sur{Zhuang}}
	\author[2]{\fnm{Yicheng} \sur{Pu}}
	\author*[1]{\fnm{Xiaolong} \sur{Wang}}\email{xiaolong@fudan.edu.cn}
	
	\affil*[1]{\orgdiv{Institute of Modern Physics}, \orgname{Fudan University}, \orgaddress{\city{Shanghai}, \postcode{200433}, \country{China}}}
	
	\affil[2]{\orgdiv{Department of Physics}, \orgname{Fudan University}, \orgaddress{\city{Shanghai}, \postcode{200433}, \country{China}}}
	
	\abstract{
		Accurate Time-of-Flight (TOF) measurement with sub-100 picosecond resolution is a critical requirement for particle identification in future high-energy physics experiments, such as the Belle II $K_{L}$ and Muon (KLM) detector upgrade. Achieving this precision with large-area Silicon Photomultipliers (SiPMs) is challenging due to the inherent junction capacitance, which degrades signal rise time. In this work, we developed and evaluated a high-time-resolution cosmic ray detector based on plastic scintillators and customized SiPM arrays. To optimize the readout for block-shaped scintillators, we systematically compared different sensor topologies. We demonstrate that a multi-face readout topology, utilizing low-capacitance 4-series (4S) SiPM modules coupled to four faces of the scintillator, achieves an excellent coincidence time resolution of approximately 68 ps, outperforming the $\sim$100 ps resolution of the concentrated 4-series 3-parallel (4S3P) hybrid topology. Furthermore, to validate the system's practical performance, we successfully measured well-known cosmic ray observables, specifically the relativistic muon velocity via TOF reconstruction. These results highlight the potential of the multi-face 4S configuration as a high-precision solution for future TOF detector upgrades.
	}
	
	
	\maketitle
	
	\section{Introduction}\label{sec1}
	
	Accurate timing measurement is a pivotal requirement for particle identification and background suppression in modern high-energy physics experiments \cite{Leo1994}. In the Belle II experiment, the endcap KLM system \cite{Abe2010} is undergoing upgrades from legacy resistive plate chambers to plastic scintillators coupled with SiPMs \cite{Aushev2015}. While SiPMs offer high photon detection efficiency, achieving sub-100 ps TOF resolution across large-area detectors introduces significant technical hurdles. A primary bottleneck for scaling up sensor areas is the cumulative junction capacitance and the inherently higher dark count rate (DCR) of large SiPM arrays. The increased capacitance severely degrades the signal rise time, while the elevated dark noise introduces continuous baseline fluctuations and signal pile-up, ultimately worsening the overall time jitters in the data acquisition (DAQ) chain.
	
	To address the capacitance issue in extended readouts, previous studies demonstrated that a 4S3P hybrid topology provides an optimal solution for long strip scintillators, successfully achieving sub-100 ps resolutions for the Belle II KLM upgrade \cite{Wang2026}. However, to manage higher particle multiplicities and provide precise spatial mapping, future high-rate experiments and specific detector regions increasingly demand highly granular TOF systems constructed from small block-shaped scintillators. While long strips are optimized for cost-effective large-area coverage, these highly segmented building blocks are essential for reducing signal pile-up and achieving superior 3D spatial resolution. Consequently, transitioning to these granular components requires a corresponding evolution in readout architecture. Simply applying the concentrated sensor topologies traditionally used for strips limits the uniformity of light collection and the ultimate timing precision of small blocks. Therefore, maximizing the performance of these highly granular systems requires dedicated systematic investigation into new multi-face sensor architectures.
	
	In this paper, we report on the development and comprehensive performance evaluation of a high-precision TOF detector system designed to optimize the readout of block-shaped plastic scintillators. We systematically compared different SiPM array configurations, specifically evaluating the concentrated 4S3P topology against a multi-face 4S readout topology. 	By decoupling the sensors into 4S modules distributed across four faces of the scintillator, the system minimizes the cumulative junction capacitance typically associated with large-area parallel branches. This symmetric configuration effectively preserves the integrity of the signal leading edge and ensures highly uniform light collection efficiency throughout the scintillator volume, thereby facilitating the achievement of sub-100 ps timing precision without the performance degradation inherent in concentrated sensor arrays. Supported by custom 3D-printed mechanical structures for stable optical coupling and a 5 GS/s high-speed digitizer, this multi-face approach achieved an intrinsic time resolution of approximately 68 ps, significantly outperforming the $\sim$100 ps resolution of the 4S3P configuration. This level of precision is highly competitive with state-of-the-art highly granular TOF systems, such as those in the MEG II and PANDA experiments, which successfully achieved sub-100 ps resolutions using small-sized scintillators \cite{Cattaneo2014MEG, Bohm2016PANDA, Nishimura2020MEG}, while specifically demonstrating the unique geometric advantages of the multi-face topology for thicker block configurations.
	
	Furthermore, to validate the system's high-precision timing capabilities in practical scenarios, the detector was deployed to measure fundamental physical properties of cosmic muons. We demonstrated its practical TOF capabilities by accurately capturing relativistic muon velocities. This comprehensive measurement provides a solid benchmark, validating the multi-face 4S configuration as a high-performance solution for future TOF upgrades.
	
	\section{Detector System and Experimental Setup}\label{sec2}
	
	\subsection{Detector Construction}
	
	The core of the detector consists of plastic scintillator blocks coupled to SiPMs. The primary light detection is performed by SiPMs manufactured by Hamamatsu Photonics K.K. In this work, the S14160-6050 SiPMs were employed due to their high photon detection efficiency, large active area, low DCR, typically $\sim$6.25 MHz, and low optical crosstalk (OCT), typically 7\% \cite{Hamamatsu_S14160}.
	
\begin{figure}[htbp]
		\centering
		\begin{minipage}{0.48\textwidth}
			\centering
			\includegraphics[width=\linewidth]{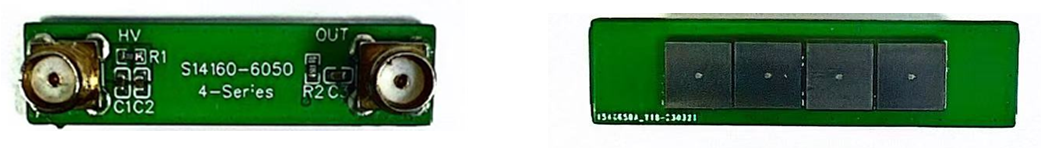}
		\end{minipage}\hfill
		\begin{minipage}{0.48\textwidth}
			\centering
			\includegraphics[width=\linewidth]{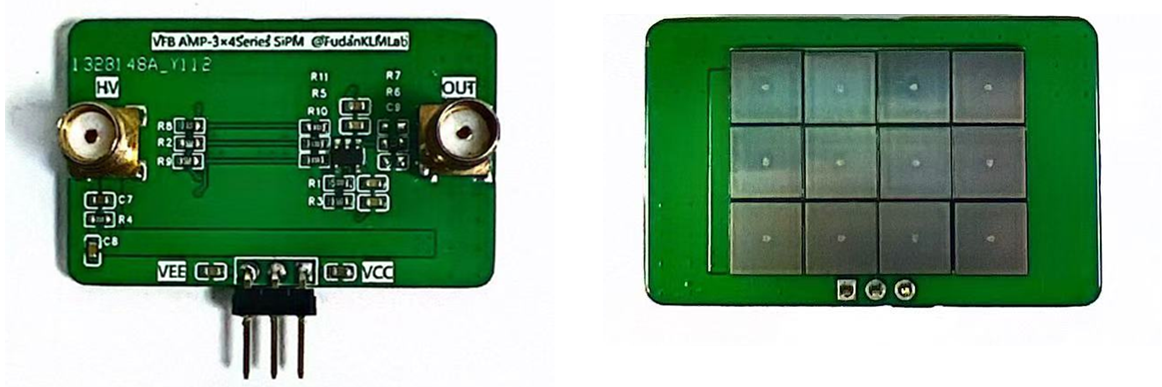}
		\end{minipage}
		\vspace{0.1cm} 
		\begin{minipage}{0.48\textwidth}
			\centerline{(a)}
		\end{minipage}\hfill
		\begin{minipage}{0.48\textwidth}
			\centerline{(b)}
		\end{minipage}
		\caption{Photographs of the custom SiPM readout boards utilized in the detector, showing both the front and back side views. (a) The 4S board, integrating four Hamamatsu S14160-6050 SiPMs connected in series. This configuration effectively mitigates input junction capacitance to significantly shorten the signal rise time \cite{Wang2026}, making it ideal for multi-face readout schemes. (b) The 4S3P hybrid board, comprising 12 SiPMs arranged in three parallel sets of four series-connected SiPMs. This hybrid configuration serves as the optimized choice for larger SiPM arrays \cite{Wang2026}, providing concentrated large-area light collection. Both configurations feature SMA connectors to ensure stable high-voltage bias input and high-fidelity signal extraction.}
		\label{fig1}
	\end{figure}

	The front-end electronics of the detector system consist of two primary stages: the SiPM sensor boards (readout boards) and the external preamplifier modules. First, the SiPMs were integrated onto custom printed circuit boards (PCBs). Two types of sensor PCB layouts were utilized: a 4S board, which connects four SiPMs in series, and a 4S3P board, comprising three parallel sets of four series-connected SiPMs, as shown in Fig. \ref{fig1}.
	
	Crucially, to maintain a stable overvoltage and ensure uniform Geiger-mode avalanche gain, the sensor boards are supported by a dedicated low-noise bias voltage circuit. Inspired by high-sensitivity avalanche photodiode power supplies, the bias circuit employs a fixed-frequency pulse-width modulation step-up converter. By operating the inductor in a discontinuous current mode and optimizing the internal switching speeds, the design inherently mitigates high-frequency voltage spikes and minimizes $di/dt$ and $dv/dt$ coupled noise. Furthermore, the predictable noise spectrum generated by the fixed switching frequency is effectively attenuated using simple LC filters. This power design provides a stable bias voltage, maintaining a low-noise baseline for the precise extraction of timing signals.
	
	To ensure the optimal electronic performance required for achieving high time resolution, the raw signals from the SiPM boards are routed to custom high-speed, low-noise preamplifiers \cite{Wang_NST_2023}. The preamplifier is characterized by a wide bandwidth of 426 MHz and a low baseline noise level of $\sigma \approx 0.6$ mV. Furthermore, a pole-zero-cancellation network is implemented in the amplification stage to significantly reduce both the rise and fall times of the SiPM signals, thereby effectively mitigating signal pile-up before digitization.
	
	Efficient optical coupling between the scintillator and the SiPM arrays was achieved by polishing the scintillator faces and applying optical silicone oil. To ensure a stable mechanical connection, custom 3D-printed connectors were designed and deployed. Furthermore, to prevent ambient light leakage and strictly minimize OCT between adjacent channels, the scintillator blocks were tightly wrapped in black light-tight electrical tape. To systematically evaluate the impact of scintillator thickness and readout topology on timing precision, three distinct detector configurations, each consisting of a pair of scintillator blocks, were assembled:
	\begin{itemize}
		\item \textbf{Configuration 1:} Two $50 \times 50 \times 10 \text{ mm}^3$ scintillator blocks, with 4S boards coupled to all four side faces.
		\item \textbf{Configuration 2:} Two $50 \times 50 \times 20 \text{ mm}^3$ scintillator blocks, with 4S boards coupled to all four side faces.
		\item \textbf{Configuration 3:} Two $50 \times 50 \times 20 \text{ mm}^3$ scintillator blocks, with 4S3P boards coupled to only two opposite side faces.
	\end{itemize}
	
	\subsection{Data Acquisition and Offline Timing Analysis}
	
	The experimental setup integrates the detector modules with a high-speed DAQ and trigger system. The plastic scintillator detectors are biased using a CAEN SY4527 high-voltage mainframe. Upon particle interaction, the initial analog signals are processed by the custom low-noise preamplifiers. To accurately capture the fast transient signals, the amplified outputs are fed into a CAEN DT5742B desktop digitizer \cite{CAEN_DT5742}. Based on the DRS4 switched-capacitor array chip, this digitizer operates at a high sampling rate of 5 GS/s with a 12-bit resolution, providing a hardware timing bin width of 200 ps. A coincidence logic unit is utilized to generate the system trigger, effectively rejecting dark noise and environmental background. The digitized waveform data is subsequently transferred to a PC via the CAEN WaveDump software for comprehensive offline analysis.
	
	Achieving the ultimate sub-100 ps time resolution relies heavily on the precision of the offline waveform processing algorithms. Traditional Leading Edge Discrimination (LED) suffers significantly from the time-walk effect, where signals of varying amplitudes cross a fixed absolute voltage threshold at different times, deteriorating the timing precision. To fundamentally eliminate this amplitude-dependent time walk, a digital Constant Fraction Discrimination (CFD) algorithm is implemented offline. The arrival time of each pulse is determined by identifying the exact point on the rising edge where the signal amplitude reaches a constant fraction, optimized at $f = 0.2$ (20\%), of its maximum peak height. 
	
	Crucially, to overcome the limitation of the 200 ps discrete sampling bins and achieve superior timing precision, a localized curve-fitting technique is applied to the signal's leading edge. Around the identified discrete threshold-crossing point, a waveform window covering $\pm 3$ sampling bins is extracted. A second-order polynomial function is then fitted to these local data points. By solving for the exact sub-bin crossing time on this continuous fitted parabolic curve, the algorithm suppresses digitization quantization errors. This waveform interpolation method ensures the precise extraction of the intrinsic time resolution limit of the SiPM arrays.
	
	\section{Measurements and Results}\label{sec3}
	
	\subsection{SiPM Characterization}
	
\begin{figure}[htbp]
		\centering
		\begin{minipage}[b]{0.48\textwidth}
			\centering
			\includegraphics[width=\linewidth]{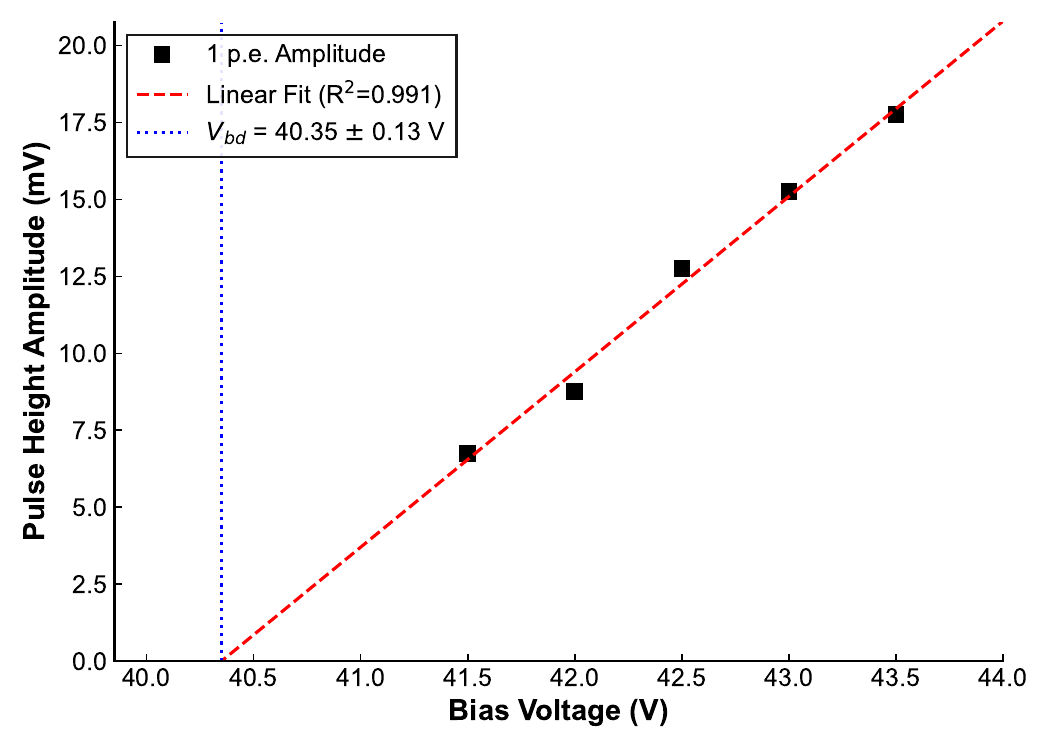} 
		\end{minipage}\hfill
		\begin{minipage}[b]{0.48\textwidth}
			\centering
			\includegraphics[width=\linewidth]{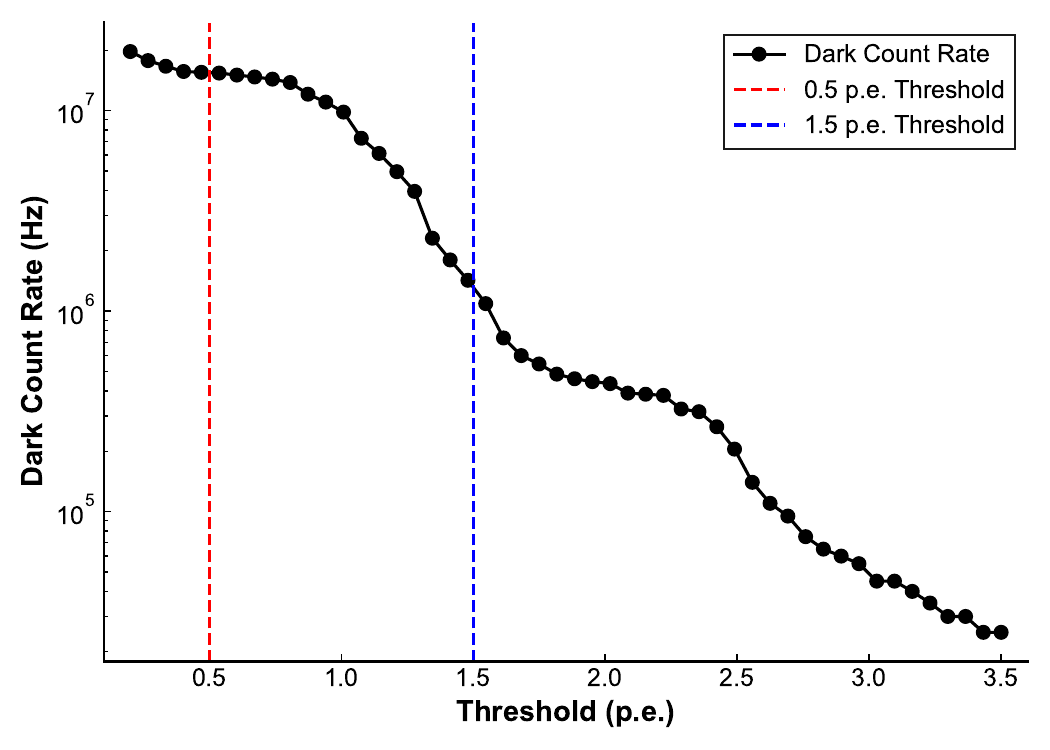} 
		\end{minipage}
		\vspace{0.1cm}
		\begin{minipage}{0.48\textwidth}
			\centerline{(a)}
		\end{minipage}\hfill
		\begin{minipage}{0.48\textwidth}
			\centerline{(b)}
		\end{minipage}
		
		\caption{SiPM characterization results in a dark environment. (a) Determination of the $V_{bd}$ using the single-photoelectron gain extrapolation method across various bias voltages. (b) DCR versus discrimination threshold scan at an operating bias of 43.0 V. The distinct plateaus demonstrate excellent single-photon resolution. The total DCR and OCT probability are extracted from the rates at 0.5 p.e. and 1.5 p.e. thresholds, indicated by dashed lines.}
		\label{fig:sipm_char}
	\end{figure}

	The Hamamatsu S14160-6050 SiPM arrays were characterized in a dark environment to establish the baseline for subsequent timing measurements. The breakdown voltage ($V_{bd}$) was determined using the gain extrapolation method, where the primary single-photoelectron peak amplitude was measured at various bias voltages. By performing a linear fit and extrapolating to zero amplitude, the $V_{bd}$ was determined to be $40.35 \pm 0.13$ V (Fig. \ref{fig:sipm_char}a).
	
	The noise characteristics were also evaluated at a nominal operating overvoltage of $+2.7$ V ($V_{bias} = 43.0$ V). To accurately quantify the noise levels and eliminate threshold-dependent ambiguity, a threshold scan was performed (Fig. \ref{fig:sipm_char}b). By extracting the event rates at the 0.5 p.e. and 1.5 p.e. discrimination thresholds, the DCR and the OCT probability were precisely determined to be $(1.54 \pm 0.03) \times 10^7$ Hz and $(7.08 \pm 0.46)\%$, respectively. These results, including their statistical uncertainties, are in good agreement with the manufacturer's specifications, confirming that the SiPMs and custom front-end electronics were operating in an stable working condition for the TOF performance evaluation.

	\subsection{Timing Performance}
	
\begin{figure}[htbp]
		\centering
		\begin{minipage}{0.32\textwidth}
			\centering
			\includegraphics[width=\linewidth]{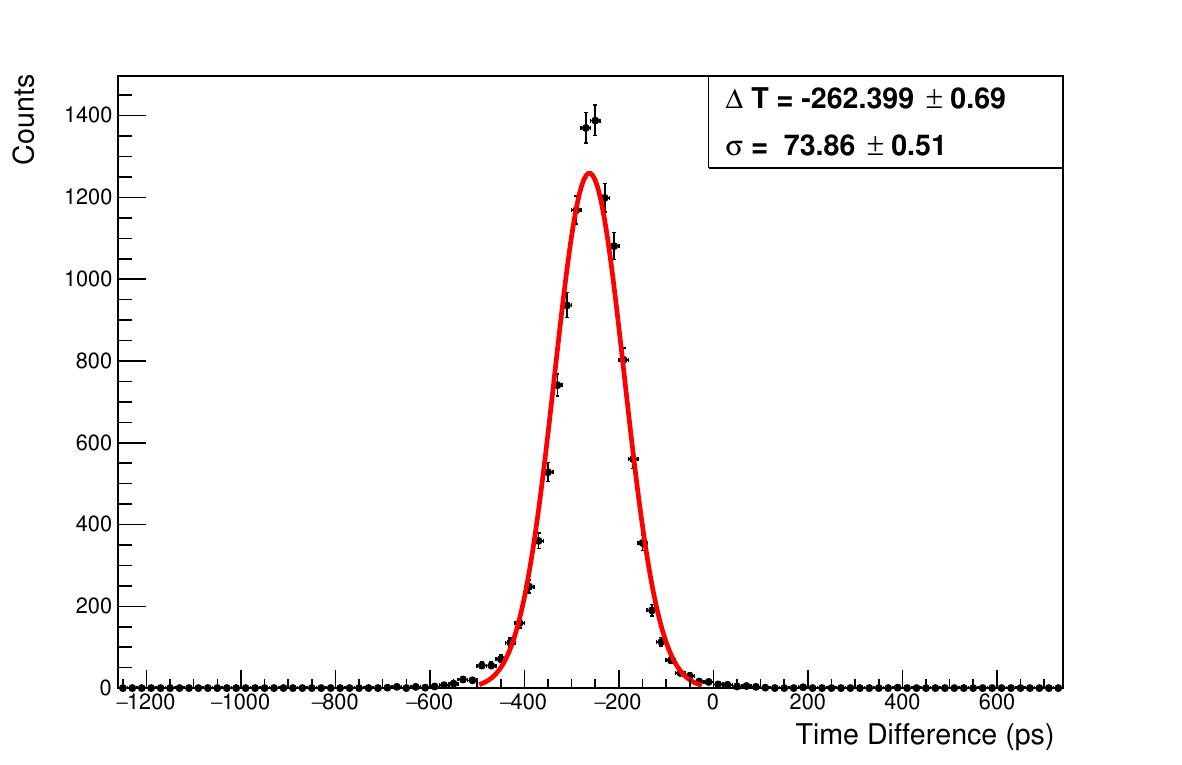}
			\caption*{(a) Config 1: 4S (10mm)}
		\end{minipage}\hfill
		\begin{minipage}{0.32\textwidth}
			\centering
			\includegraphics[width=\linewidth]{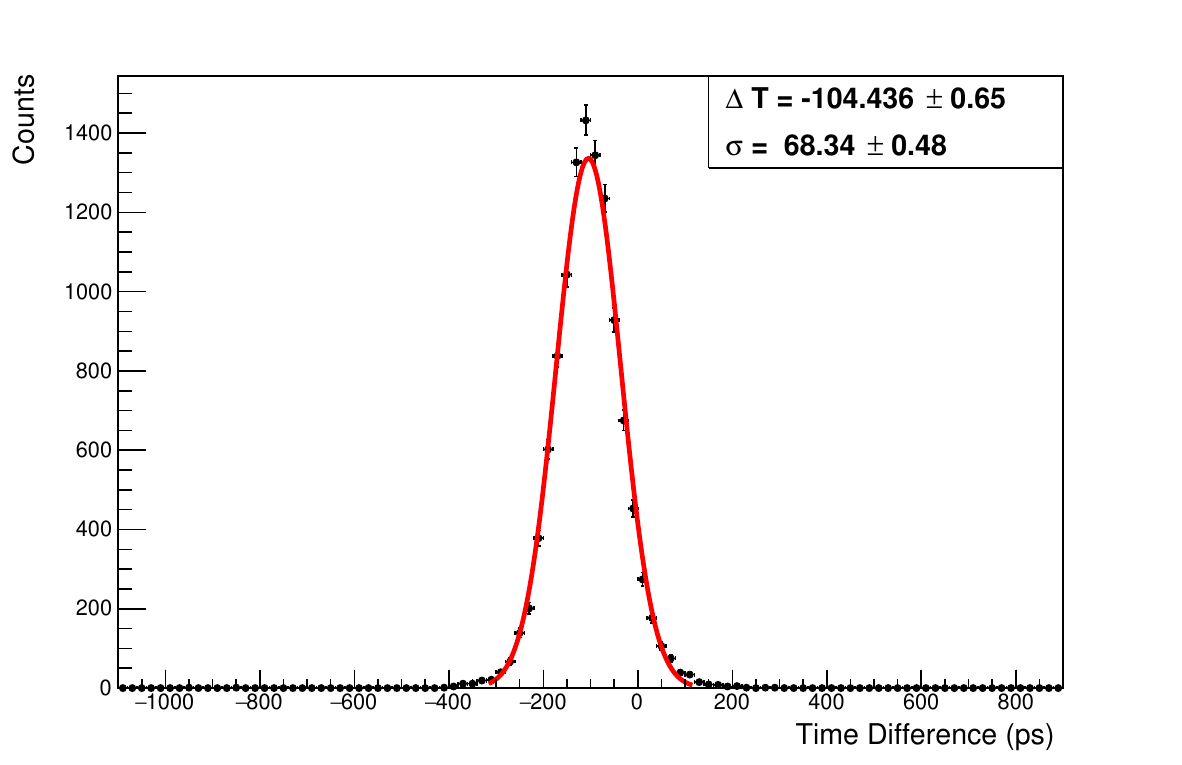}
			\caption*{(b) Config 2: 4S (20mm)}
		\end{minipage}\hfill
		\begin{minipage}{0.32\textwidth}
			\centering
			\includegraphics[width=\linewidth]{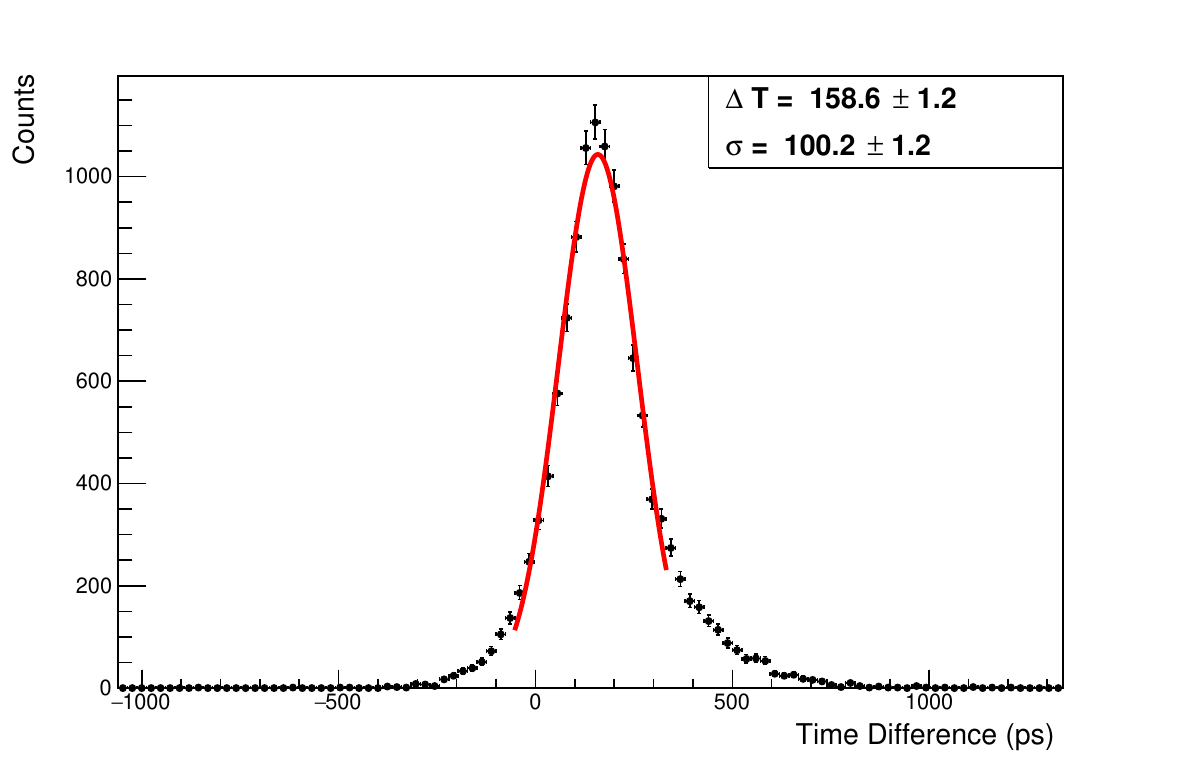}
			\caption*{(c) Config 3: 4S3P (20mm)}
		\end{minipage}
		\caption{Time resolution measurement results evaluating the three distinct detector configurations utilizing cosmic ray coincidence. The histograms represent the measured time difference distributions between paired modules, fitted with Gaussian functions to extract the coincidence time resolution standard deviation $\sigma_{\Delta T}$. (a) Configuration 1 utilizing 10 mm thick blocks yields a $\sigma_{\Delta T}$ of 73.86 ps. (b) Configuration 2 utilizing 20 mm thick blocks coupled with distributed 4S modules achieves an optimized $\sigma_{\Delta T}$ of 68.34 ps, corresponding to an intrinsic single-detector resolution of 48.32 ps. (c) Configuration 3 deploying the concentrated 4S3P hybrid topology on limited end-faces results in a broader dispersion with a $\sigma_{\Delta T}$ of 100.2 ps, illustrating the timing degradation caused by cumulative terminal capacitance.}
		\label{fig:time_resolution_results}
	\end{figure}

	\textbf{Time Resolution:} The intrinsic time resolution of the detector system was evaluated utilizing a cosmic ray coincidence setup. Two identical detector modules were vertically aligned, acting as a symmetric self-triggered coincidence system where a valid event was recorded only when coincident signals were registered by both modules simultaneously. For each recorded event, the time difference $\Delta T$ between the corresponding channels of the two detectors was calculated and histogrammed. A Gaussian function was fitted to this $\Delta T$ distribution to extract the Coincidence Time Resolution (CTR), represented by the standard deviation $\sigma_{\Delta T}$. Because this symmetric configuration utilizes the paired modules themselves for coincidence triggering without an external reference detector, the subtraction of trigger time resolution is not required. Assuming identical timing performance for both paired modules, the intrinsic single-detector time resolution $\sigma_t$ is analytically obtained by dividing the coincidence resolution by $\sqrt{2}$ (i.e., $\sigma_t = \sigma_{\Delta T} / \sqrt{2}$). As shown in Fig. \ref{fig:time_resolution_results}, the Gaussian fits yielded $\sigma_{\Delta T}$ values of 73.9 $\pm$ 0.5 ps, 68.3 $\pm$ 0.5 ps, and 100.2 $\pm$ 1.2 ps for the three tested configurations, respectively. Applying the aforementioned derivation, the corresponding intrinsic single-detector time resolutions $\sigma_t$ are determined to be 52.3 $\pm$ 0.4 ps, 48.3 $\pm$ 0.4 ps, and 70.9 $\pm$ 0.8 ps, respectively.
	
	These comparative results reveal two critical physical mechanisms governing the timing precision of large-area detectors. First, the impact of scintillator thickness is clearly demonstrated by comparing the 10 mm-thick Configuration 1 against the 20 mm-thick Configuration 2, which improved the measured CTR from 73.9 ps to 68.3 ps. In scintillation detectors, the overall timing precision is fundamentally governed by a competition between two physical effects: the statistical fluctuation of photoelectron emission, which benefits from a higher photon yield, and the optical transit time spread caused by optical path differences, which degrades as geometric dimensions increase. For minimum ionizing particles such as cosmic muons, the total energy deposition per unit path length is relatively small. Consequently, increasing the scintillator thickness from 10 mm to 20 mm provides a significantly larger energy deposition, thereby multiplying the total photon yield. Because the amplitude-dependent time-walk effect has already been corrected by the CFD timing method, this performance improvement confirms that the gain in photoelectron statistics dominates over the induced optical path dispersion within this thickness regime. Ultimately, the enhanced photon yield produces a steeper signal leading edge, minimizing the random time jitter.
	
	More importantly, comparing Configuration 2 against the dual-face 4S3P readout in Configuration 3, both utilizing a 20 mm thickness, further highlights the significant advantage of the multi-face readout architecture for block scintillators. Although Configuration 3 effectively manages junction capacitance for concentrated large arrays on limited end-faces, yielding a measured coincidence resolution of 100.2 ps, the distributed multi-face architecture in Configuration 2 outperforms it. By decoupling the total sensor area into independent, low-capacitance 4S modules distributed across four faces, this highly symmetric arrangement significantly reduces the total lumped terminal capacitance. This reduction in parasitic capacitance effectively preserves the high-frequency components of the scintillation signal and minimizes the broadening of the signal leading edge. Furthermore, the multi-face configuration ensures highly uniform light collection efficiency across the entire scintillator volume, ultimately driving the intrinsic single-detector time resolution down to the 48.3 ps level.
	
	\subsection{Cosmic Muon Velocity via TOF}
	
	\textbf{Time Calibration:} To ensure the accuracy of the subsequent velocity measurements, systematic time offsets between electronic channels ($\tau_{E2} - \tau_{E1}$) were determined using a reverse-symmetrical measurement. Specifically, an initial time difference was recorded with the detectors in a standard vertical alignment. Both detector modules were then inverted 180 degrees while maintaining the original cable connections, effectively reversing the sign of the physical flight time while keeping the electronic delays constant. By averaging the peak centers of these two measurements, the physical TOF is canceled out, allowing the systematic electronic offset to be precisely isolated. This method yielded a stable calibration constant of 64.13 ps, which was subsequently subtracted during the data analysis phase to provide the baseline correction for high-precision TOF reconstructions.
	
	To validate the high-precision timing capabilities of the multi-face 4S architecture in a practical physical application, the system was deployed to determine the average velocity of cosmic muons using the TOF method. This measurement utilizes a dual-detector timing framework analogous to our recent cosmic ray velocity measurements for the KLM upgrade \cite{Wang2026}. In this experimental configuration, two detector modules (Configuration 2) were vertically aligned and separated by a fixed baseline distance of $L = 32.5 \pm 0.1\text{ cm}$.
	
\begin{figure}[htbp]
		\centering
		\begin{minipage}{0.48\textwidth}
			\centering
			\includegraphics[width=\linewidth]{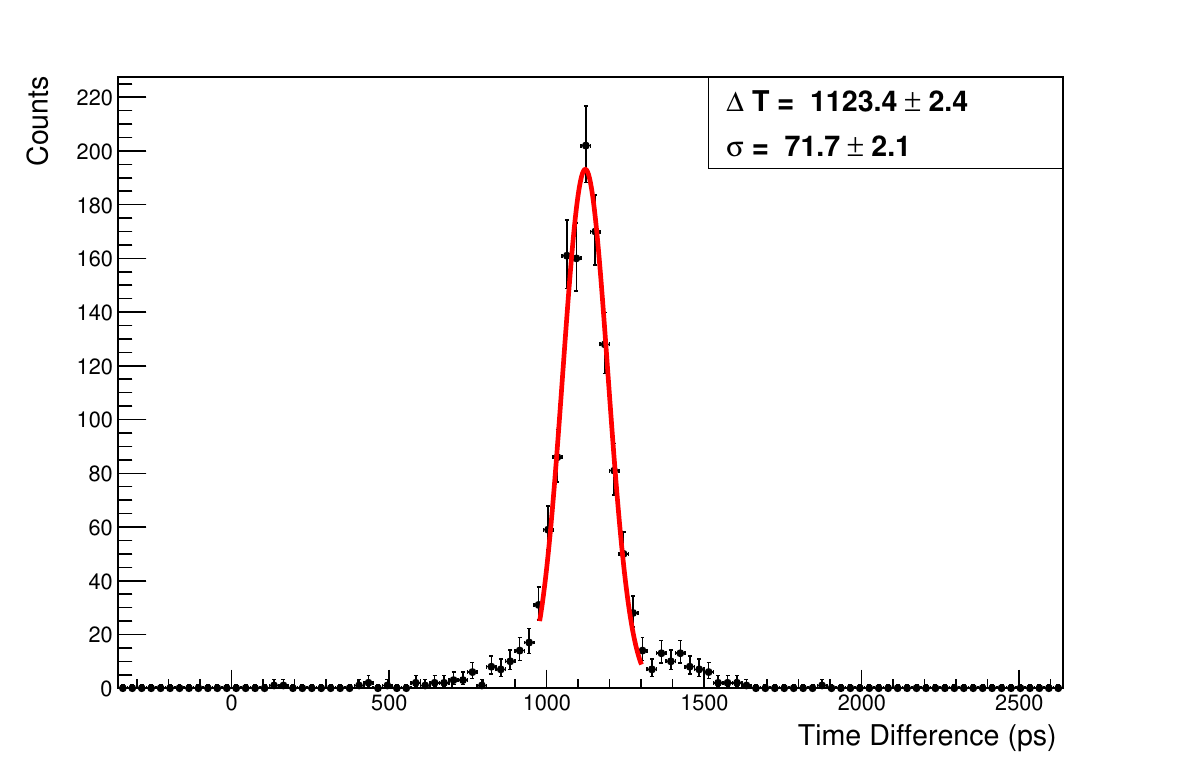}
			\caption*{(a) Calibrated Time Intervals}
		\end{minipage}\hfill
		\begin{minipage}{0.48\textwidth}
			\centering
			\includegraphics[width=\linewidth]{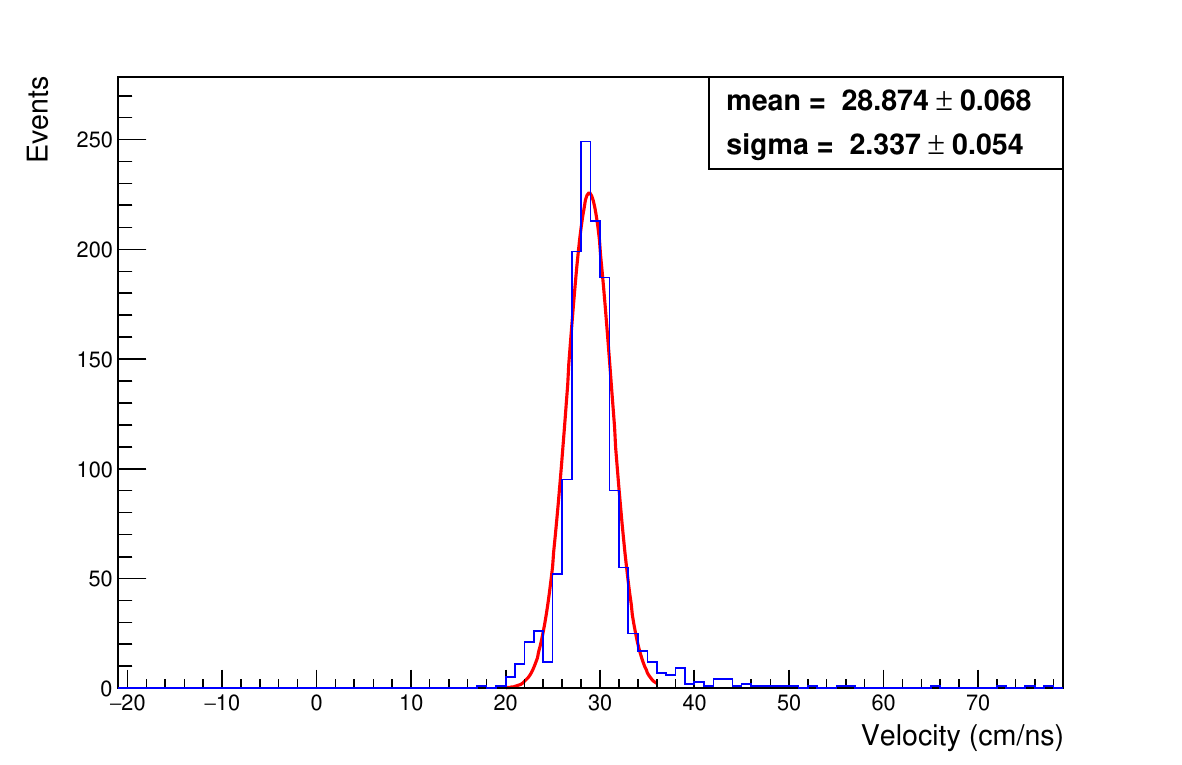}
			\caption*{(b) Muon Velocity Distribution}
		\end{minipage}
		\caption{Measurement of cosmic muon velocity using the TOF method across a fixed vertical baseline distance of 32.5 cm. (a) Distribution of the calibrated time intervals recorded between the upper and lower detector modules utilizing Configuration 2. The Gaussian fit yields a most probable flight time of 1123.4 ps with a standard deviation of 71.7 ps. Systematic electronic delays have been fully accounted for using a reverse-symmetrical calibration baseline. (b) The derived cosmic muon velocity spectrum reconstructed event-by-event from the calibrated flight times. The distribution peak corresponds to an average muon velocity of $2.887 \times 10^8\,\text{m/s}$, approximately 96.3\% of the speed of light in a vacuum, while the measured dispersion standard deviation of 2.337 cm/ns demonstrates strong quantitative self-consistency with the combined effects of instrumental timing limits and physical momentum spread.}
		\label{fig:muon_tof_velocity_results}
	\end{figure}

    Signal digitization was executed using the CAEN DT5742B digitizer at a sampling rate of 5 GS/s. The intrinsic electronic noise of the DAQ system was quantitatively evaluated utilizing pre-trigger baseline waveform segments across 1,000 recorded events, yielding a stable average baseline root-mean-square noise of 1.40 mV. Given the measured average net peak amplitude of 409.17 mV deposited by minimum ionizing cosmic muons, the multi-face readout framework maintains a signal-to-noise ratio exceeding 290. This high SNR sufficiently limits the impact of baseline noise on the signal leading edge, thereby significantly suppressing the electronic time jitter. By applying the established calibration constant to these low-jitter waveform data, systematic electronic delays and cable length offsets were strictly accounted for. The resulting distribution of the calibrated time intervals between the two detectors is shown in Fig. \ref{fig:muon_tof_velocity_results}(a).
	
	Based on the flight time measured for each muon event across the 32.5 cm baseline, the corresponding velocity distribution was generated. A Gaussian fit to this distribution (Fig. \ref{fig:muon_tof_velocity_results}(b)) yielded an average cosmic muon velocity of:
	\begin{equation}
		\bar{v}_{\text{muon}} = (2.887 \pm 0.006) \times 10^8 \,\text{m/s}
	\end{equation}
	
	This measured velocity corresponds to approximately 96.3\% of the speed of light in a vacuum, denoted as $c$, confirming the relativistic nature of cosmic muons reaching the Earth's surface. It is noted that this experimental value is slightly lower than the theoretical expectation for typical sea-level muons having an average momentum of approximately 4 GeV/$c$ and a corresponding relativistic velocity factor $\beta$ near 0.998. This systematic deviation is primarily attributed to two experimental factors. First, the finite geometric acceptance of the dual-detector setup registers coincident muons arriving at non-vertical incident angles where the zenith angle $\theta$ is greater than zero. Because the velocity reconstruction assumes a strictly vertical flight path over the baseline distance $L$ of 32.5 cm, the actual flight trajectories are systematically longer by a factor of $1/\cos\theta$, leading to an underestimation of the calculated velocity. Second, given the 20 mm thickness of the scintillator modules, slower, lower-energy muons deposit a larger amount of ionization energy per unit path length according to their specific energy loss $\text{d}E/\text{d}x$. This increased deposition results in higher photon yields and a proportionally higher trigger efficiency, thereby introducing a slight statistical bias toward the lower-momentum tail of the cosmic muon spectrum. Overall, this real-world measurement fully validates the system's sub-nanosecond timing tracking capability while faithfully reflecting the inherent geometric and physical constraints of the experimental setup.
	
	Furthermore, the measured velocity dispersion confirms quantitative self-consistency with the intrinsic timing precision of the detector framework. Based on error propagation across the fixed baseline distance $L$, the expected relative velocity resolution $\sigma_v/c$ governed purely by the CTR $\sigma_{\Delta T}$ of 68.3 ps is analytically derived as $\sigma_{\Delta T} \cdot c / L$, yielding a theoretical limit of approximately 6.30\%. The corresponding Gaussian fit to the experimental velocity spectrum exhibits a total measured standard deviation $\sigma_v$ of 2.337 cm/ns, representing a relative spread of 7.80\%. This quantitative relationship demonstrates that the instrumental timing uncertainty accounts for the dominant fraction of the observed width. The slight quadrature excess in the measured dispersion corroborates the aforementioned physical broadening mechanisms, primarily the varied geometric path lengths from non-vertical trajectories combined with the intrinsic momentum spread of the sea-level muon sample.
	
	\section{Conclusion}\label{sec4}
	
	This work demonstrates the successful implementation of an optimized readout architecture designed to meet the sub-100 ps timing criteria of upcoming high-energy physics applications, including the Belle II KLM upgrade. Addressing the timing degradation typically induced by large terminal capacitance in extensive sensor arrays, we investigated alternative coupling configurations tailored for block-shaped plastic scintillators.
	
	The experimental results highlight a clear architectural advantage: distributing independent 4S SiPM modules across four distinct faces of a 20 mm-thick scintillator yields an intrinsic single-detector time resolution of 48.3 ps (corresponding to a measured coincidence resolution of 68.3 ps). By effectively minimizing terminal parasitic capacitance, this symmetric multi-face geometry preserves the leading-edge integrity of the signal, thereby substantially outperforming the concentrated 4S3P topology conventionally deployed on limited end-faces.
	
	Furthermore, the operational capability of this optimized multi-face setup was verified through a practical physical application. Integrated with a 5 GS/s waveform digitizer, the system successfully resolved relativistic cosmic muon velocities via TOF reconstruction, demonstrating self-consistency between the derived velocity dispersion and the intrinsic detector timing limits.
	
	Ultimately, this study establishes that geometric-electronic co-optimization through distributed multi-face readouts provides a highly effective pathway for large-area timing detectors. The confirmed timing precision offers a solid quantitative foundation for implementing comparable readout topologies in future high-luminosity particle identification frameworks.
	
	\bibliography{main}
	
\end{document}